\documentclass{optica-article}

\journal{opticajournal} 

\articletype{Research Article}

\usepackage{lineno}
\usepackage{soul}
\usepackage{upgreek}

\begin{document}

\title{Incoherent Fourier Transform Spectroscopy \newline With Room-Temperature Coverage \newline From NIR To THz}

\author{Jakub Mnich\authormark{1,*}, Grzegorz Gomółka\authormark{1}, Marco Schossig\authormark{2}, Jarosław Sotor\authormark{1}, Łukasz A. Sterczewski\authormark{1}}

\address{\authormark{1}Laser and Fiber Electronics Group, Faculty of Electronics, Photonics and Microsystems, Wrocław
University of Science and Technology, Wybrzeże Stanislawa Wyspiańskiego 27, 50-370 Wrocław, Poland\\
\authormark{2}Infrasolid GmbH, Gostritzer Straße 61-67, 01217 Dresden, Germany}

\email{\authormark{*}jakub.mnich@pwr.edu.pl} 

\begin{abstract*} 
Despite the broadband nature of thermal light sources, optical spectroscopy over multiple spectral bands simultaneously remains challenging. Here, we demonstrate a practical Fourier transform infrared spectrometer (FTIR) that achieves room-temperature spectral coverage from 1~to~50~\textmu m (300--6~THz) in seconds using a single set of optics, with the long wave cutoff extendable to 90~\textmu m (3.3~THz) and the short wave to the ultraviolet (0.39~\textmu m). The interferometer employs a diamond plate beam splitter and windowless lithium tantalate (LTO) detector to probe the spectrum of combined incoherent sources operating at different temperatures. Applications of the instrument in modern chemometry, material science, and medicine are envisioned.
\end{abstract*}


\rule{\textwidth}{0.4pt}
\medskip

\noindent
Fourier transform infrared spectroscopy (FTIR) has matured into an established technique for investigating (ro)vibrational modes of chemical compounds in the near-infrared (NIR) and mid-infrared (MIR) regions. However, accessing the lower-frequency range, namely the far-infrared (FIR) and terahertz (THz) regions, remains elusive. In particular, wavelengths between 25 and 60~\textmu m (12 to 5~THz) are covered mostly with synchrotrons, specialized FTIR instruments equipped with polyethylene terephthalate (PET, Mylar) or high-resistivity silicon beam splitters and polyethylene-windowed pyroelectric detectors, or with a selection of different THz emission schemes based on air plasma \cite{dangelo_ultra-broadband_2014}, organic nonlinear crystals \cite{jazbinsekOrganicCrystalsTHz2019, sterczewski_broadband_2025} or thin-film spintronic emitters \cite{papaioannou_thz_2021, bull_spintronic_review_2021}. Unfortunately, none of these solutions is suitable for compact instruments due to their reliance on ultra-fast lasers, intensive cooling, or complex optomechanical repositioning. Wavelengths above 60~\textmu m are well served by typical antenna-based THz \cite{koch_terahertz_2023} systems, while below 25~\textmu m classic FTIRs using KBr for beam splitters and windows perform well \cite{mnich_ultra-broadband_2024}. Nevertheless, significant spectral gaps remain due to technological challenges that call for a solution. 

Advancements in laser spectroscopy techniques, especially those involving frequency combs \cite{picque_frequency_2019, coddington_dual-comb_2016}, have significantly enhanced spectroscopic capabilities for both fundamental science and gas-phase experiments. These improvements enable unprecedented resolution \cite{maslowski_surpassing_2016}, dynamic range \cite{mandon_fourier_2009, mandon_femtosecond_2007}, and data acquisition rates at the kilohertz level \cite{coddington_dual-comb_2016}. However, in the context of practical spectroscopy for non-gaseous samples—commonly encountered in the fields of chemistry, biotechnology, pharmaceutical research, etc.—the emphasis shifts towards wide spectral bandwidth rather than high resolution, given the broader absorption features characteristic of condensed-phase samples and the complexity of real-world mixtures. Since the FTIR spectral coverage cannot be matched by existing frequency combs \cite{picque_frequency_2019, maslowski_surpassing_2016}, traditional incoherent Fourier spectroscopy remains the primary workhorse for such scenarios. To extract meaningful information from measurements, modern chemometry employs multivariate analysis to limit the complexity of the dataset and isolate relevant data without having to resolve fine absorption details \cite{gredilla_non-destructive_2016}. However, this greatly benefits from multi-octave spectral coverage.

\begin{figure}[ht!]
	\centering \includegraphics{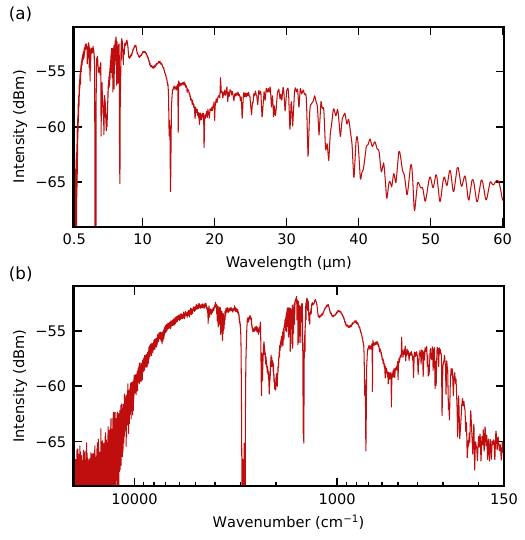} 
	\caption{Spectral coverage of the instrument in wavelength (a) and wavenumber (b) scales. The spectrum contains absorption features of H\textsubscript{2}O, CO\textsubscript{2}, and polyethylene foil windows in the flow cell. It was acquired as a single-shot with 1~cm optical path difference (OPD) and acquisition time of 10 seconds.} 
	\label{fig:coverageplot}
\end{figure}

To address this niche, in this work, we present a room-temperature, incoherent FTIR spectrometer with a single-shot spectral coverage of 1 to 50~\textmu m (6~THz) (Fig.~\ref{fig:coverageplot}a,b) operating in a static optical configuration without any actively cooled components. The range of probed wavelengths is obtained over seconds of integration time rather than minutes or hours. Although broadband coherent light sources offer superior spectral resolution and power per spectral element, the spectral coverage, robustness, and simplicity of incoherent thermal light emitters employed in the spectrometer are of practical relevance, as demonstrated here by measuring water absorption features in different wavelength ranges, all in agreement with HITRAN reference data \cite{gordon_hitran2020_2022}. We also show that with improved optics, the low-frequency coverage can extend up to 90~\textmu m (3.3~THz). Although difficult to obtain, absorption spectra in these regions contain essential features of many amino acids \cite{tan_ultra-broadband_2025} and polymers \cite{dangelo_ultra-broadband_2014} due to the collective low-frequency modes located there.  

\begin{figure}[ht!]
	\centering \includegraphics{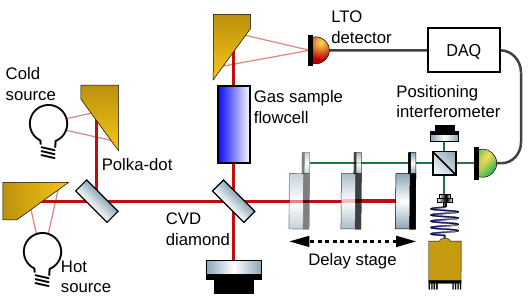} 
	\caption{Experimental setup using a combination of incoherent thermal sources operating at different temperatures. Broadband operation is possible thanks to the diamond beamsplitter and windowless LTO thermal detector. } 
	\label{fig:ftsdiagram}
\end{figure}

The experimental setup (Fig.~\ref{fig:ftsdiagram}) comprises 3 main parts: a dual-temperature thermal source, primary interferometer with a diamond beam splitter (red beam), and a reference interferometer (green beam) responsible for tracking the moving mirror position. Details of the latter are provided in Ref.~\cite{mnich_ultra-broadband_2024}.

From a source perspective, a significant limitation of thermal emitters is that their optical performance is fundamentally limited by the Planck's radiation law and the emissivity of their active surface. Increasing the temperature shifts the spectrum towards shorter (visible) wavelengths and increases the total radiated power. However, the long-wave emission tail barely improves at elevated temperatures. In principle, for a constant temperature, more radiated power can be obtained by making the emitter surface larger, albeit at the expense of worsened collimation. This is because a large-area emitter strongly deviates from a point source, which leads to a loss of spectral resolution. Regrettably, neither increasing the temperature nor making the emitter larger leads to an increased spectral dynamic range for longer wavelengths. To address this issue, we propose to combine two thermal sources operating at significantly different temperatures and then balance their spectral power densities, therefore creating the flat spectrum desired in FTIRs. A static optical configuration is preferred to opto-mechanical switching due to robustness, low complexity, and no need for spectral stitching imposed by band-selective acquisition. In implementation, collimated beams from the temperature-mismatched sources are combined on a polka-dot beam splitter employing a matrix of micro-mirrors for the beam splitting action rather than Fresnel losses. This ensures that longer-wavelength radiation can be reflected without penetrating the material, which often absorbs in this region. 

The two combined sources should operate at significantly different temperatures, so that the hot source contributes radiation close to the NIR and the cold source contributes longer wavelengths in such a way that both spectra blend seamlessly into a maximally flat power spectrum. We show that a standard quartz tungsten-halogen (QTH) is suitable as the hot source and the ceramic powder-coated (CPC) radiator (described elsewhere~\cite{mnich_ultra-broadband_2024}), maintains high emissivity up to the THz region and therefore can serve as the compact cold source. The optimal spectral region for the intercept between both sources lies around 4--6~\textmu m, where the polka-dot CaF\textsubscript{2} substrate maintains good transmission.

\begin{figure}[ht!]
	\centering \includegraphics{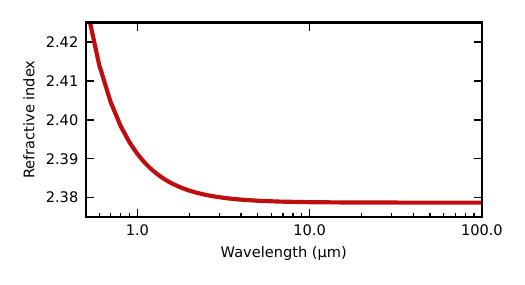} 
	\caption{Simulated diamond refractive index over wavelength range of interest as described by the Sellmeier equation \cite{turri_index_2017}.}
	\label{fig:diaxrefr}
\end{figure}

Having addressed the limitations of the source, we focus on the interferometer. Beam splitters typically used in FTIRs are based on ZnSe (1~to~14~\textmu m) or Ge-coated KBr (1.6~to~31~\textmu m) substrates and can easily limit the longest wavelength measurable by an instrument. We identified synthetic diamond as the only material that can act as a beam splitter from 1~\textmu m up to the THz range, therefore a 1", 0.5~mm-thick plate is used in the primary interferometer. Although not optimal for any spectral range in particular, reflecting only about 17\% at each interface for unpolarized light, it has almost constant optical properties from the NIR to THz, eliminating the need for optomechanical switching between multiple beam splitters. Unfortunately, below 2~\textmu m diamond's refractive index rapidly increases~\cite{turri_index_2017}, as shown in Fig.~\ref{fig:diaxrefr}. This property imposes a fundamental limitation on the short-wavelength cut-off of the spectrometer as beams from interferometer arms lose overlap at the detector and the interference signal fades. 

The last limitation results from the photodetector. Arguably, there are virtually no wavelength-agile semiconductor detectors, which requires the use of a pyroelectric materials with spectral response limited only by the absorber. The latter is known to remain effective well up to the THz range \cite{wubs_performance_2024}. While a standard deuterated triglycine sulfate (DTGS) detector can provide detectivities ($D^*$) in excess of $10^8$~Jones, it rapidly degrades in contact with moisture and therefore requires a protective window \cite{whatmore_pyroelectric_2023}. This limitation can easily be circumvented by using a high-$D^*$, thin-membrane lithium tantalate (LTO) detector capable of operating fully exposed to ambient conditions without any protective window. Such detectors are described in detail by S.~E.~Stokowski and M.~Schossig in \cite{stokowski_ion-beam_1976,schossig_broadband_2024} and with modern manufacturing techniques achieve comparable $D^* > 10^8$~Jones. For this work, a custom electronic front-end for the LTO was fabricated to allow range switching over multiple decades of sensitivity from 1~to~750~MV/A. We show that such a capability provides flexibility with regard to FTIR compatibility with different radiation sources, achieving close-to-optimal sensitivity for measurements with partially or fully enabled combined thermal sources, as well as with an externally coupled laser.

\begin{figure}[ht!]
	\centering \includegraphics{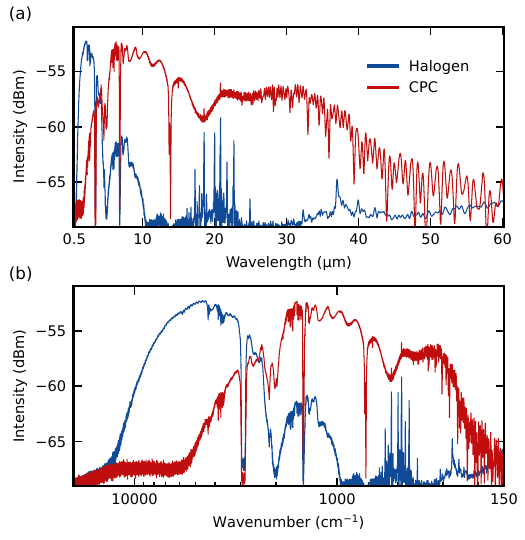} 
	\caption{Spectral regions covered by the halogen lamp and CPC sources plotted on wavelength and wavenumber scales. When powered simultaneously, they yield the spectrum in Fig.~\ref{fig:coverageplot}.}\label{fig:combinedplot}
\end{figure}

To prove the spectral coverage offered by the instrument, we resolve the absorption features of water vapor from a sample of human breath sealed in a 10 cm-long flow cell with windows made of thin polyethylene (PE) foil (Fig.~\ref{fig:ftsdiagram}). During these measurements, the instrument was sealed and purged with nitrogen to limit the ambient gas content in the beam path. Absorption dips from these PE windows are clearly visible in Fig.~\ref{fig:coverageplot} and Fig.~\ref{fig:combinedplot}, located at $\sim$725, 1466, and 2899~cm$^{-1}$ (13.79, 6.82, and 3.45~\textmu m respectively).

First, we acquired an ambient spectrum without the sample in an unpurged measurement chamber (Fig.~\ref{fig:coverageplot}), making H\textsubscript{2}O and CO\textsubscript{2} absorption lines clearly visible. The optical signal is available between 1 and 50~\textmu m with both sources active. In a combined setup, they can be enabled independently, allowing us to verify spectral contributions from both tungsten-halogen and CPC sources (Fig.~\ref{fig:combinedplot}). 

\begin{figure}[ht]
	\centering \includegraphics{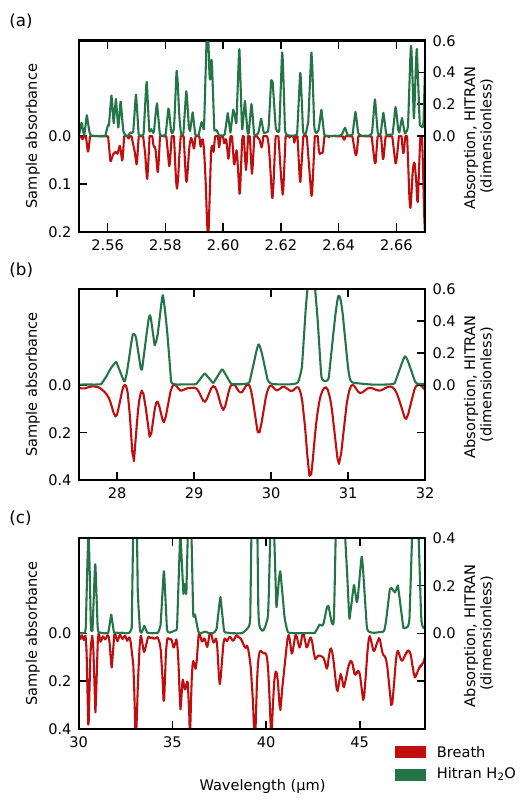} 
	\caption{Breath absorption features measured at different wavelengths and compared with HITRAN reference data for water vapor. Acquired with 100 averages, 1~cm OPD over a combined integration time of 17 minutes.} 
	\label{fig:h2oplot}
\end{figure}

Notably, the visible spikes (mainly for the halogen lamp) between 15--25~\textmu m (600--360~cm$^{-1}$) arise from vibrations of the delay line mechanism rather than being of optical origin. Acoustic interference is easily picked up by exposed pyroelectric detectors and often overlaps with the spectrum. The halogen source covers the whole visible range, as well as the NIR, but visible light does not produce interference in the setup so the spectral coverage starts at 1~\textmu m, then CPC contribution begins at 2~\textmu m and both sources intersect just before the diamond's two-phonon resonance at 4~\textmu m. During all measurements, the tungsten filament was kept at a nominal temperature of 2800~K (2530\textcelsius) and the CPC source around 870~K (600\textcelsius), although in a previous work \cite{mnich_ultra-broadband_2024} we showed that CPC sources can be effectively used at much lower temperatures if necessary. 
Despite the instrument being purged with N\textsubscript{2} for these measurements, trace amounts of H\textsubscript{2}O are still visible, especially at $\lambda$~>~30~\textmu m, where its absorption peaks are stronger. Together with diamond etalons, they dominate this region of the spectrum.

\begin{figure} [ht] 
	\centering \includegraphics{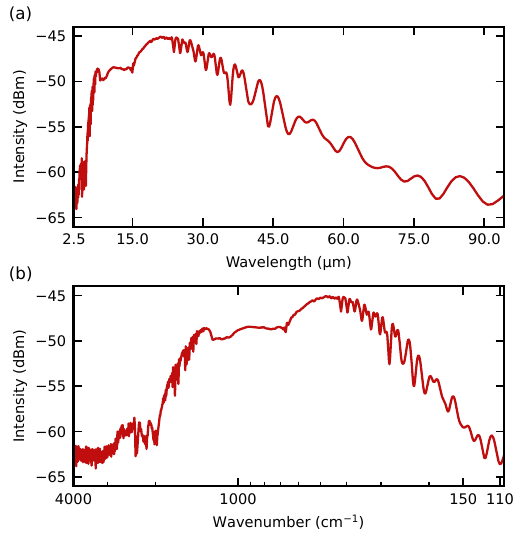} 
	\caption{Spectrum captured in a simplified configuration, with the CPC source coupled directly into the interferometer and the flowcell removed. Acquired with 30 averages, 2~mm OPD over a combined integration time of 2.5 minutes.} 
	\label{fig:globarDirect}
\end{figure}

We want to note that the amount of signal available for longer wavelengths is rather low and easily confused with acoustic noise, vibrations or the detector's own noise floor fluctuations and mechanical resonances caused by changing temperature, especially in the presence of direct heating of the pyroelectric membrane from low-interference-contrast and high-power incoherent sources. Proving spectral coverage in that region is therefore achieved by capturing water absorption features with its chamber purged with N\textsubscript{2} and a sample of human breath stored in the flow cell, as shown in Fig.~\ref{fig:h2oplot}. The measured spectrum agrees well with HITRAN reference data \cite{gordon_hitran2020_2022} using a triangular apodization window. 

H\textsubscript{2}O absorption features are well-resolved between 2.55 and 2.67~\textmu m (Fig.~\ref{fig:h2oplot}a), as well as between 28 and 32~\textmu m (Fig.~\ref{fig:h2oplot}b), where even broadband KBr beam splitter-based FTIRs struggle to deliver a useful dynamic range \cite{mnich_ultra-broadband_2024}. The setup still resolves water absorption features between 30 and 50~\textmu m (Fig.~\ref{fig:h2oplot}c), although at that point the limited dynamic range visibly distorts line shapes. 

We further show that extending coverage towards THz is made possible with improved optics. Fig.~\ref{fig:globarDirect} shows a spectrum captured with a simplified configuration, with the optical path made as short as possible, flow cell removed, source combination optics replaced with direct CPC coupling and the interferometer adjusted for maximum interference contrast at long wavelengths. Measurement is also performed with a much lower resolution (>5~cm$^{-1}$). With these modifications, the spectral coverage extends up to 90~\textmu m (3.3 THz).

\begin{figure} [ht!]
	\centering \includegraphics{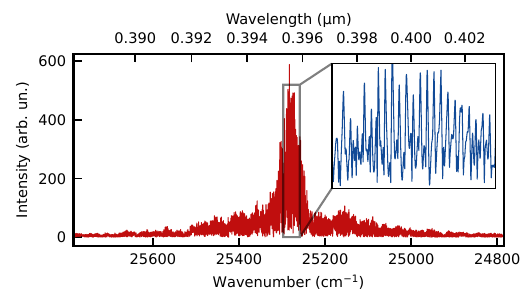} 
	\caption{Multimode UV laser diode spectrum measured by the instrument. Acquired as a single scan with a 5~cm OPD.} 
	\label{fig:uvplot}
\end{figure}

While rapid changes in the diamond's refractive index prevent incoherent sources from interfering at visible wavelengths, the interferometer can work down to the UV range with coherent sources, which we demonstrated by measuring the emission spectrum of a~multimode UV laser diode~(Fig.~\ref{fig:uvplot}). We show that the multimode laser spectrum centered around 395~nm (760~THz) is easily observed and its longitudinal modes spaced by 1.6 cm\textsuperscript{-1} (48 GHz) are clearly resolved. 

In summary, we demonstrated the broadest spectral coverage achieved in a Fourier Transform spectrometer with a static optical configuration, reaching 1--50~\textmu m. To address fundamental performance limitations of thermal radiation sources, we combined a halogen-tungsten lamp with a high-emissivity ceramic powder coated emitter (optimized for longer wavelengths) using a~CaF\textsubscript{2} polka-dot beam splitter. We proved the spectral coverage by acquiring water vapor absorption features in multiple wavelength ranges including the THz, and comparing results with the HITRAN database.

By greatly simplifying the instrument to maximize its optical efficiency, we demonstrated an extension of the spectral coverage up to 90~\textmu m (3.3~THz), which shows that, with improved optical throughput and solutions presented in this work, spectrometers that cover more than 6 octaves in single-shot measurements at room-temperature and with a static optical configuration are technically viable. Although rapid changes in the diamond beam splitter's refractive index prevent it from working in the visible range, we experimentally demonstrated that the same broadband interferometer used for other experiments still measures reasonably good spectra of spatially coherent sources even in the UV range. A~multimode laser diode with an emission peak at 395~nm has clearly resolved longitudinal modes spaced by 1.6~cm\textsuperscript{-1} when measured by our instrument. Such ultra-broadband spectrometers can operate with low tens of watts of electrical power and cm-scale beam paths, making them promising candidates for future miniaturization and use in modern multivariate chemometry methods, which become more selective and precise as more information-carrying data points are made available.

\medskip

\begin{backmatter}
\bmsection{Funding}
J. M., G. M., and Ł. A. S. acknowledge funding from the European Union (ERC Starting Grant, TeraERC, 101117433). Views and opinions expressed are, however, those of the authors only and do not necessarily reflect those of the European Union or the European Research Council Executive Agency. Neither the European Union nor the granting authority can be held responsible for them.

\bmsection{Acknowledgment}
LTO crystals used in custom pyroelectric detectors were delivered by Laser Components Germany~GmbH and DIAS Infrared~GmbH.

\bmsection{Disclosures}
Marco Schossig is an employee of Infrasolid GmbH, the manufacturer of the prototype CPC thermal source used in this work.

\bmsection{Data Availability Statement}
Data underlying the results presented in this paper are available in Ref.~\cite{NBFLBZ_2026}.

\end{backmatter}

\bibliography{literature}

\end{document}